\documentclass[12pt,A4]{article}
\usepackage[margin=1in,footskip=0.25in]{geometry}

\usepackage{graphicx, color}  
\usepackage{amsmath,amssymb, amsfonts, MnSymbol, mathrsfs}
\usepackage{natbib}
\usepackage{hyperref}

\begin{document}

\title{Model Validation Practice in Banking: A Structured Approach for Predictive Models}
\author{Agus Sudjianto, H2O.ai and University of North Carolina at Charlotte\\ Aijun Zhang, Wells Fargo\thanks{The views expressed in this paper are those of the authors and do not necessarily reflect those of Wells Fargo.}} 
\date{Revision: 10/15/2024}
\maketitle

\begin{abstract}%
This paper presents a comprehensive overview of model validation practices and advancement in the banking industry based on the experience of managing Model Risk Management (MRM) since the inception of regulatory guidance SR11-7/OCC11-12 over a decade ago. Model validation in banking is a crucial process designed to ensure that predictive models, which are often used for credit risk, fraud detection, and capital planning, operate reliably and meet regulatory standards. This practice ensures that models are conceptually sound, produce valid outcomes, and are consistently monitored over time. Model validation in banking is a multi-faceted process with three key components: {\bf conceptual soundness evaluation}, {\bf outcome analysis}, and {\bf ongoing monitoring} to ensure that the models are not only designed correctly but also perform reliably and consistently in real-world environments. Effective validation helps banks mitigate risks, meet regulatory requirements, and maintain trust in the models that underpin critical business decisions.

\end{abstract}

\section{Introduction}  In the modern banking landscape, models play pivotal roles in critical decision-making processes, from credit risk assessment and fraud detection to capital planning and stress testing. As these models become more sophisticated and their impact on financial institutions grows, the need for robust validation practices has never been more pressing. Model validation in banking is not merely a regulatory checkbox; it is a crucial process that ensures the reliability, accuracy, and effectiveness of the models upon which banks heavily rely.

{\bf Model risk} refers to the potential for adverse consequences resulting from decisions based on incorrect or misused models. These risks can lead to significant financial loss, poor decision-making, and reputational damage for banks. Model risk arises primarily for two reasons: (1) a model may contain fundamental errors and produce inaccurate outputs, and (2) even a correct model may be used inappropriately, especially when applied in conditions beyond its design scope, as indicated by the regulatory guidance SR11-7/OCC11-12 \citep{SR11-7}. Managing model risk begins with understanding the business purpose of the models, identifying what can go wrong, and considering unintended consequences. These potential downsides need to be translated into measurable metrics within the model, for which a threshold of acceptable risk can be set.

Model validation, therefore, should not only focus on ensuring that a model is technically sound but also on identifying potential root causes of failure. This includes quantifying the impact of these failures on the business and finding ways to either prevent or mitigate the associated risks. The ultimate goal is to ensure that models are both accurate and resilient in dynamic environments.

The complexity of financial markets, coupled with the rapid advancement of machine learning and artificial intelligence techniques, has led to the development of increasingly intricate models. While these models offer unprecedented predictive power, they also introduce new challenges in terms of interpretability, reliability, and potential for unintended consequences. In this context, model validation serves as a critical safeguard, helping to identify and mitigate risks associated with model use.

 Though we refer to this process as ``validation'', it should ideally begin during model development. According to SR11-7/OCC11-12, “as a practical matter, some validation work may be most effectively done by model developers and users'' \citep{SR11-7}.
It highlights that early validation activities by developers are essential to ensuring that models are conceptually sound and properly implemented. By integrating validation steps during development, issues can be detected and addressed before the model is fully deployed, reducing the potential for downstream problems. Importantly, even though model developers might participate in early validation steps, these should always be subject to critical review by independent parties to maintain objectivity and ensure a thorough validation process.  

This paper provides a comprehensive approach for predictive model validation. The depth and rigor of the validation steps outlined in this paper should be commensurate with the model's risk level. Models with higher risk warrant a more comprehensive validation process, covering all aspects in greater detail, while lower-risk models may require emphasis on certain elements, such as basic outcome analysis and periodic monitoring. 

We aim to provide a review and advancement of model validation practices in banking since the inception of regulatory guidance on Model Risk Management (MRM) through SR11-7/OCC11-12 over a decade ago, focusing on three key components: conceptual soundness evaluation, outcome analysis, and ongoing monitoring. We explore the methodologies, challenges, and best practices associated with each of these components, drawing on both recent advancements and practical industry experiences.

The conceptual soundness evaluation section delves into the foundational aspects of model validation, including data quality assessment, input design, and model structure evaluation. Ensuring that models are built on solid theoretical grounds and align with established financial and statistical principles is crucial.

In the outcome analysis section, we explore techniques for identifying model weaknesses, assessing the reliability of model outputs, and evaluating model robustness against input noise and environmental changes. This section emphasizes rigorous testing and analysis to ensure models perform as intended across a range of scenarios.

The ongoing monitoring section addresses the critical need for continuous oversight of model performance. Strategies for detecting data drift, concept drift, and other factors that may impact model effectiveness over time are discussed. Additionally, we explore the importance of periodic testing and revalidation to ensure models remain compliant with regulatory requirements and business objectives.

Throughout the paper, we emphasize the interdisciplinary nature of model validation, which combines elements of statistics, business, computer science, and regulatory compliance. By providing a comprehensive overview of model validation practices, this paper aims to serve as a valuable resource for banking professionals, risk managers, regulators, and researchers working at the intersection of modeling and technology.

\section{Conceptual Soundness Evaluation}

{\bf Conceptual soundness} refers to assessing the foundation and logical underpinnings of the data and model, ensuring they align with the intended use case and industry standards. This stage involves evaluating the model design, data and inputs, assumptions, and optimization strategies. The objective is to confirm that the model is theoretically appropriate, interpretable, and capable of producing reliable results.

\subsection{Data Quality and Suitability}
Ensuring data quality and suitability is a foundational aspect of model validation in banking. High-quality data is essential for building reliable predictive models, especially in high-stakes environments where decisions based on model outputs can significantly impact customers and financial performance. This component focuses on verifying that the data used for modeling is accurate, complete, relevant, and appropriately processed.


\subsubsection{Data Completeness} 
The dataset should be complete, with no missing values or incomplete records. Missing values can introduce bias or lead to inaccurate predictions if not handled properly. For example, in a credit risk model, missing income data can significantly affect the assessment of a borrower’s risk profile.

\subsubsection*{Data Checks for Missing Values:}
\begin{itemize}
\item {\sf Identifying Missing Data:} Regularly assess the dataset for missing values using techniques like summary statistics or visualizations (e.g., heatmaps) to identify patterns of missingness.

\item {\sf Handling Missing Values:} Implement strategies for addressing missing data, which may include:
\begin{itemize}
\item {\sf Imputation:} Filling in missing values using methods like mean, median, or mode imputation, or more sophisticated techniques like $K$-nearest neighbors or regression imputation.
\item {\sf Deletion:} Removing records with missing values if the proportion is small and won’t significantly impact the dataset.
\item {\sf Categorical Encoding:} For categorical variables, treating missing values as a separate category can sometimes help retain useful information without distorting the dataset.
\end{itemize}
\end{itemize}

\subsubsection{Data Accuracy} 
The data should be accurate and correctly reflect the true values of the underlying variables. Inaccurate data can lead to poor model performance and misinformed decisions. For example, errors in recording customer credit scores can severely impact a model designed to assess credit risk.

\subsubsection*{Data Checks for Accuracy:}
\begin{itemize}
\item {\sf Cross-Validation Against External Sources:} Whenever possible, validate the data against trusted external datasets or known benchmarks to ensure accuracy.
\item {\sf Automated Validation Rules:} Implement rules to flag anomalies, such as credit scores outside realistic ranges (e.g., negative scores or excessively high scores) for review.
\end{itemize}

\subsubsection{Data Consistency} 
Data should be consistent across the dataset and should adhere to the same formats and standards. Inconsistencies can arise from data entry errors, differing measurement units, or variations in categorical values.

\subsubsection*{Data Checks for Consistency:}
\begin{itemize}
\item {\sf Standardization of Formats:} Ensure all data entries conform to a specified format (e.g., date formats, currency symbols).
\item {\sf Data Type Checks:} Implement checks to verify that numeric fields contain only numerical data, categorical fields have defined categories, and dates are valid.
\end{itemize}

\subsubsection{Data Relevance} 
The features used in the model must be relevant to the problem being addressed. Irrelevant features can introduce noise and lead to overfitting, negatively impacting model performance.

\subsubsection*{Data Checks for Relevance:}
\begin{itemize}
\item {\sf Domain Knowledge Consultation:} Involve subject matter experts to determine which variables are likely to have predictive power for the task at hand.
\item {\sf Feature Importance Analysis:} After initial modeling, analyze feature importance to identify which features contribute meaningfully to the model’s predictions.
\end{itemize}

\subsubsection{Data Outliers} 
Outliers are extreme values that differ significantly from other observations in the dataset. They can arise from measurement errors, data entry mistakes, or actual variability in the data. In banking, outliers can significantly affect model performance, especially in models sensitive to extreme values (e.g., linear regression).

\subsubsection*{Data Checks for Outliers:}
There are various {\bf statistical methods} for checking data outliers \citep{zhao2019pyod}, including 
\begin{itemize}
\item {\sf Univariate Checks:} Basic techniques such as $Z$-score or the IQR (Interquantile Range) method can be used to identify univariate outliers. Values that fall outside 1.5 times the IQR above the third quantile or below the first quantile can be flagged for further investigation.
\item {\sf Isolation Forest:} This is an advanced outlier detection method that works by isolating observations in a dataset \citep{liu2008isolation}. It creates random partitions and identifies anomalies based on how quickly they can be isolated from the rest of the data. Isolation Forest is particularly effective for high-dimensional datasets and can help identify outliers that traditional univariate methods might miss.
\item {\sf DBSCAN (Density-Based Spatial Clustering of Applications with Noise):} This clustering method identifies outliers based on the density of data points in the feature space \citep{schubert2017dbscan}. Points in low-density regions compared to the surrounding data points are classified as outliers.
\item {\sf Local Outlier Factor (LOF):} This technique measures the local density deviation of a given data point with respect to its neighbors, helping to identify points that have a significantly lower density than their neighbors, thus flagging them as potential outliers \citep{breunig2000lof}.
\item {\sf PCA with Mahalanobis Distance}: this combined approach leverages the strengths of both techniques to enhance outlier detection.
\begin{itemize}
\item[a)] Principal Component Analysis (PCA) is first applied to reduce the dimensionality of the dataset while preserving its variance. This helps to visualize and capture the primary structure of the data, making it easier to identify outliers.

\item[b)] Once the data is transformed into the principal component space, the Mahalanobis distance is calculated for each observation.  Mahalanobis distance measures how far a point is from the mean of a distribution, taking into account the covariance among variables. Points with a large Mahalanobis distance relative to a defined threshold (e.g., based on the chi-squared distribution) are flagged as potential outliers.
\end{itemize}
This combined method is particularly effective in high-dimensional datasets, as PCA helps mitigate the curse of dimensionality, while Mahalanobis distance allows for an understanding of the multivariate relationships among features.
\end{itemize}

Other than statistical methods, one may also check data outliers by {\bf visualization techniques}, e.g. employing box plots or scatter plots to visually inspect the data for potential outliers. This helps in understanding the distribution and identifying unusual observations.
 
Once outliers are identified, perform {\bf contextual examination} by examine the outliers in context to determine if they are valid observations (e.g., high-income individuals) or if they result from errors or anomalies (e.g., data entry mistakes).

After identifying outliers, one must decide on appropriate handling strategies, such as:
\begin{itemize}
\item {\bf Capping:} Transforming outlier values to a maximum or minimum threshold to mitigate their influence on model training.
\item {\bf Exclusion:} Removing outliers from the dataset if they are determined to be erroneous or not representative of the population being modeled.
\end{itemize}

\subsection{Input Design and Control}
{\bf Input Design and Control} is an essential part of model validation that focuses on selecting high-quality, relevant features (variables) and ensuring that the input data is appropriate for the model. This step involves addressing feature selection, feature engineering, and controlling for data quality to ensure that the model functions optimally.

\subsubsection{Feature Engineering}
Feature engineering is the process of transforming raw data into meaningful input variables (features) that can improve the performance of machine learning models. It involves selecting, creating, or modifying features to make patterns in the data more evident for the model, leading to better predictive accuracy. The quality of the features directly impacts the model’s performance, interpretability, and robustness.

Constructing interpretable features -- those that are easy to understand and have a clear relationship with the outcome of interest -- is important to ensure that models are not only highly predictive, but also conceptually sound in the following sense.
\begin{itemize}
\item {\sf Improved Interpretability:} Interpretable features directly contribute to making the model more transparent, as they provide a clear understanding of how each input affects the prediction. This is particularly important in regulated industries where decision transparency is essential.
\item {\sf Model Simplicity:} By using well-designed, interpretable features, you reduce the need for complex model structures. This can result in simpler models that perform well without relying on black-box techniques, making the overall system more understandable and easier to maintain.
\item {\sf Trust and Explainability:} Models built on interpretable features are easier to explain to stakeholders, customers, and regulators. This improves trust in the model's decisions, as the reasoning behind predictions is clear.
\item {\sf Regularization and Robustness:} Interpretable features often lead to more robust models by avoiding overfitting. Models that depend on simple, meaningful features generalize better to new data and are less sensitive to noise or irrelevant information.
\end{itemize}

When designing features for interpretability, it’s important to:
\begin{itemize}
\item {\sf Simplify Complex Variables:} Break down complex or multi-dimensional data into simpler, intuitive components. For example, instead of using a raw time series, you might extract specific trends or seasonal components that are easier to interpret.
\item {\sf Use Domain Knowledge:} Use knowledge from the specific domain to create features that align with business logic or scientific understanding. For example, in finance, constructing features like debt-to-income ratio or credit utilization provides more meaningful insight than raw financial data alone.
\item {\sf Avoid Overly Complex Interactions:} While interactions between features can improve model performance, they should be chosen carefully to ensure they remain interpretable. Simple interactions, such as the product of two features with clear relationships, can be more interpretable than complex polynomial combinations.
\item {\sf Consider Aggregation and Binning:} Create features by aggregating data points (e.g., averages, sums, or counts) or by binning continuous variables into categorical ranges. For example, transforming a continuous variable like age into bins (e.g., "18-25", "26-35", etc.) can make the feature more interpretable while still retaining its predictive power.
\item {\sf Consider Monotonic Features:} Features that have a monotonic relationship with the target variable (where increases or decreases in the feature consistently lead to increases or decreases in the outcome) are more interpretable. For example, higher values of a feature like income generally lead to a better credit score, and this clear trend aids in model understanding.
\end{itemize}

Feature engineering plays a critical role in model design by ensuring that the input features are both interpretable and relevant. Constructing clear, meaningful features enhances model transparency, performance, and the ability to explain decisions, making them particularly valuable in high-stakes and regulated environments.

\subsubsection{Embedding Approaches}
Embedding is the process of transforming raw data into structured representations, making it suitable for machine learning tasks like classification, regression, and clustering. This transformation can occur in either lower-dimensional or higher-dimensional spaces, depending on the complexity of the data and the relationships between features.

In traditional statistical modeling, similar transformations are commonly referred to as variable transformations or basis function expansions. These approaches, such as polynomial expansions or logarithmic transformations, have long been used to represent non-linear relationships in a linear form. Embeddings, however, extend these concepts by leveraging more advanced techniques, enabling machine learning models to capture richer patterns and interactions in the data.

This section focuses on various embedding approaches that project data into useful representations, exploring both lower-dimensional and higher-dimensional transformations. These embeddings are essential for improving the performance of downstream machine learning tasks by creating representations that are more structured, manageable, and interpretable.

\begin{enumerate}
\item {\sf Principal Component Analysis (PCA) and Factor Analysis: Lower-Dimensional Embeddings.}  PCA and Factor Analysis are classical dimensionality reduction techniques that create lower-dimensional embeddings by identifying the most important components or latent factors in the data. These methods reduce noise and redundancy by projecting the data onto a smaller set of dimensions while retaining the most informative features.

Specifically, PCA projects high-dimensional data onto orthogonal axes that account for the most variance in the data. Factor Analysis, on the other hand, identifies latent variables that explain correlations among observed features.
 

\item {\sf Neural Networks: Flexible Projections to Lower and Higher Dimensions.}
Neural networks provide highly flexible embedding frameworks that can generate either lower-dimensional or higher-dimensional embeddings, depending on the architecture. Neural networks are particularly adept at capturing complex relationships and non-linear interactions in data.

\begin{itemize}
    \item {\sf Lower-Dimensional Embeddings:} Autoencoders are a type of unsupervised neural network that compress input data into a smaller set of latent variables by passing it through a bottleneck layer. This lower-dimensional embedding reduces noise and focuses on the most essential patterns in the data.
 
    For example, in a customer transaction dataset, an autoencoder might reduce hundreds of purchase behaviors into a smaller latent representation that preserves the key patterns for predicting customer churn.
 
    \item {\sf Higher-Dimensional Embeddings:} Multilayer Perceptrons (MLPs) or attention mechanisms can generate higher-dimensional embeddings by applying non-linear transformations across multiple layers. These embeddings are useful when the original data contains complex feature interactions that are difficult to capture with simpler models.
 
    For example,  an MLP could project features like credit history and transaction behavior into a higher-dimensional space, allowing for more nuanced classification of credit risk.
\end{itemize}
\item {\sf Tree-Based Models: Higher-Dimensional Embeddings.} Tree-based models, such as Random Forests and Gradient Boosted Trees, create higher-dimensional embeddings by encoding complex, non-linear feature interactions. These models do not explicitly reduce dimensionality but instead create richer representations by expanding the input space into decision paths or leaf indices.

Specifically, tree-based models generate embeddings by mapping data into new feature spaces where decision rules, paths, or leaf indices represent interactions between input variables.
 
The embedding outcome is a higher-dimensional representation of the data, capturing complex, non-linear relationships that simpler models may not detect, thus enabling more accurate prediction models. It creates one-hot-encoding for downstream task of regression or classification. See \citep{cui2023enhancing} for such an example of reformulating gradient boosted decision tree as an generalized linear model of the terminal nodes. 


\item {\sf Kernel Methods in Support Vector Machines (SVMs): Projections to Higher Dimensions.} Kernel methods in SVMs create higher-dimensional embeddings by projecting data into spaces where complex, non-linear relationships become linearly separable. This is achieved using kernel functions, which allow the model to operate in a higher-dimensional feature space without explicitly computing the transformation.

The common kernels like the Radial Basis Function (RBF) or polynomial kernels map input data into higher-dimensional spaces, allowing models to find linear decision boundaries more easily.
 
The embedding outcome is a higher-dimensional embedding where complex patterns in the data are more separable, improving the performance of tasks like classification or regression.
 
For example, an RBF kernel applied to a transaction data could transform variables into a higher-dimensional space, enabling the SVM to predict fraudulent transactions with greater accuracy.

\item {\sf Simple Polynomial Embeddings: Capturing Non-Linear Interactions.} Polynomial embeddings are a straightforward yet powerful technique for generating higher-dimensional embeddings by expanding input features into polynomial terms, including interaction terms, squared terms, or higher-order terms. These terms capture non-linear relationships between the original features, transforming the data into a space where models can learn more complex patterns, thus 
improving model performance.
 
The embedding outcome is a higher-dimensional feature space where non-linear relationships are more apparent and can be modeled more easily.
 
For example, in a real estate dataset, a polynomial embedding could expand features like square footage and number of rooms into higher-order terms (e.g., (square footage)$^2$, (square footage) $\times$ (number of rooms)), helping a regression model capture more complex relationships between house size and price.

\item {\sf Fourier and Basis Function Expansions: Embeddings for Periodic and Non-Linear Patterns.} Fourier and basis function expansions are commonly used for both lower- and higher-dimensional embeddings. These transformations are especially effective at capturing periodic and non-linear patterns in the data. Specifically, the Fourier transformations project data into a frequency domain to capture periodic patterns, while  the basis function expansions use functions like polynomials or splines to capture complex non-linear relationships.
 
The embedding outcome is a compact frequency-based embedding (Fourier) or an expanded feature space (basis function expansions) that enhances the ability of models to capture periodic or non-linear interactions.
 
For example, Fourier embeddings are useful for time-series data like weather forecasts, where periodic patterns (e.g., temperature fluctuations) are crucial for making accurate predictions. Polynomial basis function expansions, on the other hand, are ideal for capturing non-linear interactions in datasets with complex feature relationships.

\item  {\sf Attention Mechanisms: Higher-Dimensional Embeddings with Contextual Focus.} Attention mechanisms, particularly from transformer models, can generate higher-dimensional embeddings that focus on the most relevant features for a given task. This method assigns different importance weights to features, creating embeddings that capture contextual dependencies.

 
The embedding outcome is a context-dependent embedding that emphasizes the most relevant features and relationships for the task at hand.
 
For example, in a loan approval system, an attention-based model might generate embeddings that prioritize income and debt-to-income ratio for some applicants, while focusing on payment history for others, depending on the context.

\end{enumerate}

The selection of an appropriate embedding technique depends on the dimensionality of the original data and the complexity of the relationships between features:
\begin{itemize}

\item Lower-Dimensional Embeddings (e.g., PCA, autoencoders) are suited for cases where the data is noisy, redundant, or overly complex. These embeddings simplify the data while retaining key patterns, making it easier for downstream models to learn from the data.
 
\item Higher-Dimensional Embeddings (e.g., tree-based models, polynomial expansions, attention mechanisms) are useful when the data contains non-linear relationships that need to be modeled in a richer, expanded feature space. These embeddings capture complex patterns and interactions that cannot be easily represented in the original space.
\end{itemize}

Embedding methods are essential for preparing data by transforming it into structured, lower- or higher-dimensional representations. Lower-dimensional embeddings simplify the data, removing noise and redundancy, while higher-dimensional embeddings expand the feature space to capture complex, non-linear relationships. Techniques like neural networks, tree ensembles, kernel methods, and polynomial expansions provide flexible options for generating useful embeddings, each tailored to the specific needs of the data and the downstream task.

\subsubsection{Variable or Feature Selection}
The goal of feature selection is to identify the most relevant and influential variables while excluding irrelevant or redundant features. This process helps improve model performance, prevent overfitting, and enhance interpretability. Proper feature selection is particularly important in banking models where interpretability is critical for decision-making and regulatory compliance. To select the most appropriate variables, several techniques can be employed:
\begin{itemize}
\item {\sf Filter Methods:} Statistical techniques like correlation analysis, mutual information, or variance thresholding are used to identify and exclude weak predictors that do not contribute meaningfully to the outcome.
\item {\sf Wrapper Methods:} Methods such as recursive feature elimination (RFE) evaluate different combinations of features and rank them based on their impact on model performance. These techniques help to fine-tune the selection of features based on their predictive power.
\item {\sf Embedded Methods:} These methods, such as LASSO (Least Absolute Shrinkage and Selection Operator) or Ridge regression \citep{hastie2015statistical}, perform feature selection during the training process by adding regularization penalties to reduce the impact of irrelevant variables. This ensures that only the most important features contribute to the final model.
\item {\sf Conditional Independence Tests for Causality:} In banking, where causality is crucial for making decisions (e.g., determining whether a customer’s income directly affects loan default risk), conditional independence tests can be used to identify causal relationships between variables. Conditional independence tests assess whether a feature (e.g., income) is conditionally independent of the target variable (e.g., default) given other variables (e.g., credit history). Features that are not conditionally independent are more likely to have a causal relationship with the target variable and are prioritized during feature selection.
\item {\sf Causal feature selection} is particularly valuable in regulatory environments, as it helps banks understand the drivers behind model predictions. It also enhances model robustness by ensuring that the selected features have a direct impact on the outcome, rather than being merely correlated with it. For example, conditional independence tests might reveal that while a customer’s geographical location is correlated with default risk, it may not be a causal factor when income and credit history are considered.
\end{itemize}

\subsubsection{Input Control}
Managing and controlling the quality of inputs is essential for ensuring that the model is not influenced by noise, irrelevant features, or biased data. Banks must continuously monitor the inputs fed into the model to detect shifts or anomalies in the data distribution. Sensitivity analysis is often used to assess how variations in input features affect model predictions. For example, a small change in a customer’s income should not cause a dramatic shift in their creditworthiness unless other factors (such as debt levels) are also altered significantly.

Additionally, data validation checks can help ensure that the inputs remain stable and valid over time. For example, in a credit scoring model, monitoring the consistency of income and employment status data over time helps maintain the accuracy and reliability of the model.

\subsection{Model Design, Methodology Selection, and Assumptions}
{\bf Model Design and Methodology:} The design and methodology of the model must be suitable for the business objective and the type of data it will process. In banking, models are frequently used for tasks such as credit scoring, fraud detection, loss forecasting/stress testing or loan approval. It is essential to choose a modeling approach that aligns with the structure of the data (e.g., logistic regression for binary classification problems like default prediction, or decision trees for explainability). More advanced models, such as neural networks or ensemble methods, may be selected for complex tasks where traditional statistical models may fall short. However, model complexity must be justified, particularly in a regulatory context where interpretability is critical. {\sf Inherently interpretable machine learning} is highly relevant because they offer the best of both worlds—high performance while maintaining interpretability (see Section 2.4). Unlike black-box models, these models allow users to directly understand the relationships between input features and predictions, making it easier to trust, explain, and debug decisions. 

{\bf Model Assumptions:} Every model relies on assumptions about the data and relationships between variables. These assumptions need to be clearly stated and validated against the actual characteristics of the data. For instance, linear regression models assume linear relationships between inputs and outputs, while credit risk models may assume independence between predictors. Failure to validate these assumptions could lead to unreliable model performance, especially under changing economic conditions. 

{\bf Benchmarking:} A critical aspect of conceptual soundness is comparing the performance of the model against alternative models or established benchmarks. This process involves testing the model against simpler or more interpretable models (see Section 2.4 Explainability and Interpretability) to ensure that any increase in complexity leads to a demonstrable improvement in predictive power or interpretability. Benchmarking can also involve comparison against industry-standard models or regulatory expectations.

Benchmarking with both more complex models and simpler, inherently interpretable models is essential for achieving a balance between performance and interpretability.
\begin{itemize}
\item {\sf Complex models} like deep learning or ensemble methods can capture intricate patterns and interactions in data, potentially improving prediction accuracy. Benchmarking with these models helps explore whether there is a significant performance gain compared to simpler models and justifies their use when higher accuracy is crucial, such as in highly competitive or high-stakes scenarios.
\item {\sf Simpler models} such as decision trees, linear models, or generalized additive models (GAMs) provide inherent interpretability, making it easier to understand and explain the results. Benchmarking with simpler models ensures that, whenever possible, an interpretable model can be chosen if it achieves similar performance to more complex alternatives. This is particularly important for applications where transparency, regulatory compliance, and trust in the model's decisions are required.
\item {\sf Inherently interpretable models}  (see Section 2.4) allow organizations to achieve high accuracy without sacrificing transparency, which is crucial in regulated industries like finance and healthcare. Inherently interpretable models provide a reliable, efficient, and transparent solution, avoiding the need for post-hoc explanation techniques while still delivering high-quality predictions.
\end{itemize}

{\bf Sound Modeling Practices:} Ensuring that the selected modeling approach aligns with sound statistical and machine learning practices is crucial. For example, in credit risk models, there is a regulatory expectation that the models follow the principles of Basel frameworks. Such practices include robust out-of-sample testing, model validation on unseen data, and evaluating model performance under various economic scenarios. Moreover, internal benchmarks and peer comparisons can be applied to validate the appropriateness of the model and ensure it adheres to the highest industry standards.

\subsection{Explainability and Interpretability}
{\bf Explainability:} The model explainability is crucial, particularly in high-stakes decision-making environments like banking, where financial decisions directly affect customers and regulatory compliance. For example, decision trees and logistic regression models are inherently more interpretable because they provide direct insight into how predictions are made. However, more complex models like neural networks or ensemble methods often require explainability tools such as Partial Dependent Plot (PDP; \cite{friedman2001greedy}), Accumulated Local Effect (ALE; \cite{apley2020visualizing}) or local explainability techniques such as Shapley Additive Explanation (SHAP; \cite{lundberg2017shap}) or Local Interpretable Model-agnostic Explanations (LIME; \cite{ribeiro2016lime}) to ensure that the decision-making process is understood and can be communicated clearly to stakeholders, including regulators. However, these post-hoc explainability tools are approximations; thus, they may not accurately explain the model. Complex machine learning can also be made inherently interpretable when their architectures are properly constrained as described below. See also \cite{yang2020enhancing} for a constructive approach to explainable neural networks through interpretability constraints and \cite{sudjianto2021design} for a practical guide of developing inherently interpretable machine learning models. 

\subsubsection{Locally Interpretable Machine Learning Models}
There are available locally interpretable machine learning models, such as ReLU Deep Neural Networks (DNNs) and Boosted Linear Trees, allow for clear, region-specific explanations of their predictions. In ReLU DNNs, the model behaves as a piecewise linear function, where each input point falls into a region defined by ReLU activations. Similarly, in Boosted Linear Trees, each terminal node contains a linear model, and predictions are made by routing inputs to specific leaves, where the local linear model governs the prediction. This provides exact local interpretability without needing post-hoc explanations. 

\subsubsection*{Deep ReLU Networks}
A Deep ReLU Network is a type of deep neural network that uses the Rectified Linear Unit (ReLU) as its activation function, defined as $\sigma(x) = \max(0, x)$, which outputs the input directly if positive and returns zero otherwise. This activation introduces non-linearity into the network, enabling it to model complex relationships in the data.

In a deep ReLU network, each layer applies a linear transformation followed by the ReLU activation, forming a hierarchical structure that learns increasingly abstract features from the input. 

A ReLU network is locally interpretable because it acts as a piecewise linear function. The network divides the input space into regions, each defined by a specific activation pattern, where it behaves as a local linear model. For any input, the network predictions are governed by a corresponding local linear model, providing exact local interpretability. Therefore, there is no need for post-hoc explanation methods like LIME or SHAP, which approximate local behaviors. See details in \citep{sudjianto2020unwrapping}. 

\subsubsection*{Boosted Linear Trees}
A Boosted Linear Tree model, like in LightGBM \citep{ke2017lightgbm}, is a decision tree where each terminal/leaf node contains a linear model instead of a constant value. The tree partitions the data, and within each terminal node, a linear model is fitted to the data points that fall into that node. This approach combines the non-linear partitioning power of decision trees with the predictive strength of linear models within each segment of the data.

The model is locally interpretable because each input follows a path to a specific terminal node, where a local linear model is applied. Additionally, the linear models from different terminal nodes can be aggregated, and the aggregation of linear models results in another linear model. Therefore, this structure not only enables local interpretability but also preserves the overall linear relationship in the data across segments. Thus, it provides exact local explanations, making it easier to understand the model's behavior without the need for post-hoc explanation techniques.

\subsubsection{Functional ANOVA Structure for Globally Interpretable Model}
The functional ANOVA (fANOVA) structure can be used to constrain machine learning models to be globally interpretable by breaking down the model into main effects and low-order interactions. This decomposition simplifies the model by focusing on the most important components and interactions, allowing users to understand how individual features and their combinations influence the model's predictions. Because fANOVA models focus on capturing low-order interactions and main effects, they offer a natural framework for global interpretability, ensuring the model is not only predictive but also understandable at a high level, without the need for post-hoc explanations.

The fANOVA structure is a method that decomposes a complex function (or model) into simpler components, specifically main effects of individual features and low-order interactions between features. This allows the model to capture how each feature individually affects the output and how certain combinations of features contribute to predictions.
In fANOVA, the function $f(x)$ is expressed as a sum of additive components:
$$
f(x) = g_0 + \sum_j g_0(x_j)  + \sum_{j<l} g_{jl}(x_j, x_l) + \cdots 
$$
where $g_0$ is the overall mean, $g_{j}(x_j)$'s are the main effects of individual features, and $g_{jl}(x_j, x_l)$'s capture pairwise interactions between features. Higher-order interactions can also be included, but typically only low-order interactions (e.g., pairwise) are considered for interpretability.

There are several different implementation of fANOVA models with second-order interactions \citep{lou2013accurate, yang2021gami, hu2023interpretable}. Typically, the model construction steps involve the following:
\begin{itemize}
\item {\sf Decomposition:} The model function is decomposed into main effects and interaction terms, ensuring that the complexity is manageable and interpretable.
\item {\sf Regularization:} To maintain interpretability, regularization techniques can be applied to limit the complexity of interactions (focusing on a few low-order interactions) and emphasize the importance of main effects.
\item {\sf Machine Learning:} Machine learning models, such as gradient boosting or neural networks, can be trained to estimate these components. The learning process identifies the most important features and interactions, ensuring the model is expressive while remaining interpretable.
\end{itemize}

{\sf Global Interpretability:} By focusing on main effects and low-order interactions, the resulting model is globally interpretable, meaning the behavior of the model across the entire input space is understandable. Each feature's contribution and interaction can be explicitly understood without the need for complex post-hoc explanation techniques. This fANOVA approach constrains the model complexity and ensures interpretability while still leveraging the power of machine learning to find patterns and interactions in the data.

{\sf Regulatory Compliance:} Banks must comply with regulatory requirements for model transparency and risk management. In jurisdictions where models are subject to regulatory review or approval, such as under SR 11-7 guidelines \citep{SR11-7}, explainability is essential. CFPB (Consumer Financial Protection Bureau) Circular 2022-03 \citep{CFPB2022-03} clarifies that creditors using complex algorithms for credit decisions must comply with the Equal Credit Opportunity Act (ECOA) by providing specific reasons for any adverse actions taken against applicants, such as credit denials. Regardless of the technology used, creditors are required to disclose precise reasons related to the decision-making process and cannot excuse noncompliance by claiming that their algorithms are too opaque to understand. Creditors must also ensure the accuracy of any post-hoc explanations, as such approximations may not be viable with less interpretable models. The model validation process must provide clear justifications for the model’s structure, inputs, and outputs, ensuring that it meets legal and regulatory standards.

\subsection{Parameter and Hyperparameter Optimization}
{\bf Model Parameters:} The parameters of a model are the coefficients or values that are learned during the training process. For example, in a linear regression model, the parameters are the weights associated with each input feature. These parameters must be estimated correctly using well-established techniques such as maximum likelihood estimation or gradient-based optimization to ensure accurate predictions.

{\bf Hyperparameter Tuning:} Hyperparameters, unlike model parameters, are set before training and control the learning process. Examples include the regularization strength in a logistic regression model or the number of layers in a neural network. In banking models, hyperparameter tuning is crucial to avoid both underfitting and overfitting. Techniques like grid search or random search, often combined with cross-validation, are employed to find the optimal hyperparameter values that balance model complexity and performance. Regularization techniques (e.g., L1 or L2 penalties) may be applied to prevent overfitting, especially when dealing with high-dimensional financial data.

Model replication and  stability testing are essential components of parameter and hyperparameter choice assessment.
\begin{itemize}
\item {\bf Model Replication:} Replicating the model involves building it anew using different samples of data or subsets (e.g. via bootstrapping) to verify that it produces consistent results. This helps validate the model's performance across various datasets and ensures that its predictions are not merely artifacts of specific training data.
\item {\bf Stability Testing:} Stability testing assesses whether the model's predictions remain consistent over time and across different segments of the population. Key aspects include:
\begin{itemize}
\item {\sf Random Seed Variation:} Stability testing should involve using different random seeds during the model training and testing split. By varying the random seed, banks can evaluate how changes in the data partitioning affect model performance. If the model yields similar performance metrics across different seeds, it suggests that the model is stable and not overly sensitive to specific data splits.
\item {\sf Stochastic Optimization Initialization:} Many machine learning models utilize stochastic optimization methods (e.g., stochastic gradient descent). The initialization of parameters and the random sampling of data points can influence the final model performance. By running the model with different random seeds for initialization, banks can assess whether the model converges to similar solutions consistently. Significant variations in model performance due to different initializations may indicate instability and the need for further investigation or adjustments.
\end{itemize}
\end{itemize}

\section{Outcome Analysis}
{\bf Outcome analysis} is a critical component of model validation in banking, designed to assess the performance and behavior of the model in real-world applications. This analysis aims to determine how well the model's predictions align with actual outcomes and whether the model remains reliable and accurate under various conditions. Outcome analysis is essential for identifying any deficiencies in the model, improving its robustness, and ensuring that it adapts to changes in input data and environmental conditions. In banking, this is especially important as models are used for high-stakes decisions like credit scoring, fraud detection, and risk management. Outcome analysis process focuses on four key components: identifying model weaknesses, assessing the reliability of the model outputs, evaluating robustness against input noise to prevent benign overfitting, and testing the model’s resilience to distribution drift and environmental change. Among others, the PiML toolbox \citep{sudjianto2023piml} provides a suite of model diagnostic tools for outcome analysis. 

\subsection{Identification of Model Weakness}
The first step in outcome analysis is to systematically identify weaknesses or shortcomings in the model. This involves evaluating the model’s performance under a wide range of conditions and use cases to uncover areas where it may struggle or produce unreliable results. In banking, this is especially important as models are used for high-stakes decisions like credit scoring, fraud detection, and risk management. Key methods for identifying model weaknesses include:
\begin{itemize}
\item {\sf Performance Decomposition:} Decomposing the model's performance across different segments of data, such as geographic regions, or loan categories, helps identify whether the model performs well across all cases or struggles with certain subgroups. For example, a credit scoring model may perform well overall but exhibit higher error rates for minority applicants, indicating a potential fairness issue.
\item {\sf Error Analysis:} Examining the types of errors the model makes (e.g., false positives, false negatives) can reveal specific conditions under which the model fails. For instance, a loan approval model might falsely predict low-risk customers as high-risk, leading to missed lending opportunities.
\item {\sf Backtesting and Stress Testing:} Regular backtesting (comparing model predictions with actual historical data) and stress testing under extreme conditions (e.g., financial crises or market shocks) are useful for detecting weaknesses that only emerge under particular economic scenarios.
\end{itemize} 

Identifying these weaknesses early allows the bank to address vulnerabilities before the model is deployed in high-stakes environments, such as credit approval or risk management processes.

A primary goal of model weakness identification is to identify segments or clusters where the model is weaker and prone to errors. Understanding these weaknesses helps improve the model and ensures that it remains reliable across different population groups, data ranges, or scenarios. Weakness identification typically involves performance decomposition, identifying key variables and ranges of values that contribute most to the model’s underperformance, and determining where the model overfits or underfits.

\subsubsection*{Performance Decomposition by Segments or Clusters}
Performance decomposition involves breaking down the model’s performance across different subgroups or clusters of the population to pinpoint where it performs well and where it struggles. These subgroups could be based on geographic regions, loan types, or other relevant dimensions in the banking context.
\begin{itemize}
\item {\sf Segmentation by Key Variables:} The model’s predictions are analyzed across various subgroups based on key variables like loan type, loan-to-value, and credit score. For instance, a credit risk model might perform well for middle-income borrowers but poorly for high-income or low-income groups. By segmenting the data, banks can identify specific areas where the model underperforms.
\item {\sf Clustering for Latent Patterns:} Clustering techniques such as k-means clustering or hierarchical clustering can be used to group similar instances together based on input features, without pre-defined segments. This allows the identification of latent patterns in the data where the model’s performance varies significantly. For example, a cluster of borrowers with thin credit history and low credit scores might exhibit high model error rates, indicating a potential model weakness in handling high-risk borrowers.
\end{itemize}

\subsubsection*{Identifying Variables and Their Value Ranges Contributing to Weakness}
Once performance decomposition has identified underperforming segments or clusters, it’s important to investigate the specific variables and ranges of values that are most associated with model weakness. This step helps pinpoint why the model struggles in certain areas and provides actionable insights for improvement.
\begin{itemize}
\item {\sf Variable Importance Analysis:} After segmenting the data, a variable importance analysis is conducted to identify which features (e.g., income, debt-to-income ratio, loan term) contribute the most to prediction errors. For example, in a credit scoring model, the analysis might reveal that income and credit history have the largest impact on prediction errors in certain population segments, such as high-net-worth individuals. This suggests that the model may need more refined handling of income variability or credit history for such segments.
\item {\sf Range Analysis:} Once important variables are identified, their specific value ranges contributing to model weakness are further analyzed. For example:
\begin{itemize}
\item Loan-to-Value (LTV) Range: A model might perform well for borrowers with low LTV range but produce errors for high LTV. This range analysis helps identify where the model struggles to generalize, leading to inaccurate risk assessments for extreme income values.
\item Credit Score Range: The model may have high accuracy for credit scores between 600 and 750 but fail to predict accurately for individuals with scores below 500 or above 800, where the risk factors may behave differently and require additional model refinement.
\end{itemize}
By identifying these specific value ranges, model adjustments can be made to improve its performance in weak areas. For instance, creating additional feature interactions or non-linear terms for high-income borrowers could reduce errors in that segment.
\end{itemize}

\subsubsection*{Underfitting and Overfitting Detection}
Another critical aspect of identifying model weakness is determining whether the model is underfitting or overfitting in specific segments or for certain features. Understanding these behaviors ensures that the model remains generalizable and reliable.

{\bf Underfitting:} Underfitting occurs when the model is too simple to capture the underlying patterns in the data, resulting in poor performance across all or some segments. In performance decomposition, underfitting is identified by examining areas where the model consistently makes errors or where its predictions are far from the actual outcomes. Common signs of underfitting include:
\begin{itemize}
\item {\sf High Error Rates Across Segments:} If certain segments, like borrowers with high LTV and debt-to-income, consistently show higher error rates, the model might be underfitting in those areas due to a lack of complexity or missing interactions between variables.
\item {\sf Bias in Predictions:} If the model tends to produce overly simplified predictions (e.g., always predicting low risk for all borrowers in a particular segment), this can indicate underfitting. Adding more features, introducing nonlinear or interaction terms, or using a more sophisticated model might be required to capture the nuances in these segments.
\end{itemize}

{\bf Overfitting:} Overfitting occurs when the model becomes too complex and fits the noise in the training data, leading to poor generalization to new, unseen data. Overfitting can often be identified through:
\begin{itemize}
\item {\sf Excessively Low Training Errors but High Test Errors:} In certain segments, the model may show near-perfect performance during training (indicating overfitting) but fail to generalize well on new data from the same segment. For example, the model may overfit to high-income borrowers by capturing specific patterns in the training data that do not generalize to other high-income borrowers.
\item {\sf Overly Complex Patterns for Small Segments:} Overfitting can occur when the model learns overly complex or irrelevant patterns for small or rare segments of the data. For instance, a loan approval model may overfit to a small group of very high-risk borrowers, leading to poor predictions for similar but distinct cases in the future.
\end{itemize}

{\bf Regularization techniques} such as L1/L2 regularization, dropout layers, or early stopping can help prevent overfitting by controlling model complexity and ensuring that the model generalizes well across all segments.

\subsubsection*{Actionable Insights for Model Improvement}
Identifying segments with underfitting or overfitting provides actionable insights for model refinement. This might involve:
\begin{itemize}
\item {\sf Adding Interaction Terms:} For segments where the model is underfitting, adding interaction terms between variables can help capture more complex relationships (e.g., interaction between income and employment history).
\item {\sf Regularization:} To prevent overfitting in certain segments, regularization techniques can be applied to simplify the model and reduce the impact of irrelevant variables.
\item {\sf Segment-Specific Models:} In some cases, creating separate models for different population segments (e.g., separate credit risk models for high-net-worth individuals and low-income borrowers) may improve overall performance.
\end{itemize}

\subsection{Reliability or Output Uncertainty}
{\bf Prediction uncertainty} refers to the lack of confidence in a model’s predictions, which can arise from factors such as noisy data, data sparsity, or model limitations. Identifying regions with high uncertainty is crucial because it highlights where the model's predictions are less reliable, allowing for targeted improvements. Factors driving this uncertainty include data quality issues (e.g., noise, outliers, or shifts in distribution) and instability in model predictions due to feature interactions or poor fit in specific regions.

One of the core objectives of outcome analysis is to assess the {\bf reliability} of the model predictions and estimate the {\bf uncertainty} associated with its outputs. In banking, understanding and managing the uncertainty of a model is critical to avoid undue risks in decision-making processes.
\begin{itemize}
\item {\sf Uncertainty Estimation:} Models inherently produce predictions with varying degrees of uncertainty. For instance, in a credit risk model, uncertainty may arise from incomplete data or outlier behavior. Techniques such as conformal prediction can be used to quantify prediction intervals, providing banks with a confidence level around each prediction.
\item {\sf Prediction Reliability:} Banks must evaluate whether the model predictions are consistent and reliable over time. Reliability is especially important in high-stakes decision-making, where erratic model behavior could lead to significant financial losses. Calibration plots (comparing predicted probabilities to actual outcomes) and Brier scores (measuring prediction accuracy) and conformal prediction with quantile regression are often used to evaluate reliability.
\item {\sf Quantifying Risk of Model Misuse:} Banks should also analyze the reliability of the model in scenarios of model misuse or misinterpretation, ensuring that decision-makers understand when the model's outputs are less trustworthy and how to act on them responsibly. For example, uncertainty in creditworthiness predictions may lead to more conservative loan approvals under uncertain conditions.
\end{itemize}

To address these issues and improve the model, consider:
\begin{itemize}
\item {\sf Uncertainty Quantification (UQ):} Use methods like conformal prediction to generate prediction intervals that offer a measure of confidence for each prediction. Split conformal prediction is a method for uncertainty quantification that generates prediction intervals with statistical guarantees \citep{vovk2005, shafer2008conformal}. The approach is model-agnostics, simple and computationally efficient, offering a way to estimate uncertainty for any machine learning model.

{\bf Key Steps in Split Conformal Prediction:}
\begin{enumerate}
\item Data Splitting: The dataset is split into two parts—one for training the model and the other for calibration. The model is trained on the training set.
\item Residual Calculation: The model’s residuals (differences between actual values and predictions) are calculated on the calibration set.
\item Prediction Interval Construction: Using the residuals from the calibration set, a prediction interval is constructed for each new prediction. This interval contains the true outcome with a certain probability, providing a measure of uncertainty.
\end{enumerate}
\item {\sf Feature Sensitivity Analysis:} Identify which features contribute most to uncertainty and focus on improving data quality or feature engineering for those.
\item {\sf Model Selection, Regularization and Retraining:} Consider alternative modeling approach, apply techniques like regularization, hyperparameter tuning, and noise-resistant training methods to make the model more robust in regions of high uncertainty.
\end{itemize}

By assessing output uncertainty, banks can ensure that decision-making is based on sound probabilities and mitigate the risk of unforeseen losses due to overly optimistic or pessimistic predictions.

\subsection{Robustness against Input Noise to Avoid Benign Overfitting}
A robust model should remain reliable even when exposed to small changes or noise in input data. {\bf Benign overfitting}, which occurs when a model fits noise or minor variations in the training data, is a common problem in complex models, particularly in machine learning. In banking, benign overfitting can lead to models making accurate predictions on historical data but failing to generalize to new, unseen data. 
This can result in poor decision-making in dynamic environments.
\begin{itemize}
\item {\sf Noise Sensitivity Testing:} To assess robustness, small perturbations are introduced into the input data to evaluate how much the model’s predictions change. For instance, a small change in a customer’s credit score or income level should not result in dramatically different loan approval outcomes. Sensitivity analyses, which introduce noise in input features, help ensure that the model is not overfitting to insignificant variations.
\item {\sf Invariance Testing:} Another approach is to test for invariance—the model should produce the same output even when irrelevant or redundant features are altered. For example, slight changes in non-critical inputs (e.g., formatting changes in application data) should not affect the model’s predictions.
\item {\sf Regularization Techniques:} Models should be built with regularization techniques (e.g., L2 regularization, dropout layers) that constrain model complexity and prevent overfitting. This reduces the model’s reliance on noise in the data and ensures that it generalizes well to unseen examples.
\end{itemize}

By focusing on robustness against noise, banks can prevent the harmful effects of benign overfitting and ensure that their models perform reliably on new data in real-world conditions.

To improve the robustness of machine learning models, e.g. Gradient-Boosted Decision Trees (GBDT) with XGBoost implementation \citep{chen2016xgboost}, it is essential to first identify the factors driving the model's sensitivity to noise. The key factors contributing to this sensitivity include:
\begin{itemize}
\item {\sf Overfitting:} Complex models tend to overfit to training data, making them sensitive to small perturbations in new or unseen data.
\item {\sf Feature Interactions:} Non-linear interactions between features can amplify noise sensitivity, particularly if irrelevant or weakly correlated features are heavily weighted.
\item {\sf High Variance in Decision Trees:} Individual decision trees in GBDT may be too specific to the training data, leading to instability when applied to new samples.
\item {\sf Outliers:} Outliers in the training data can disproportionately influence model performance, especially if they are not properly handled.
\item {\sf Unstable Input Features:} Features with high variance or noisy data can cause the model to produce unreliable predictions.
\end{itemize}

The techniques for improving model robustness include
\begin{enumerate}
\item {\sf Regularization:} Apply L1 (Lasso) or L2 (Ridge) regularization to penalize model complexity, reducing the weight of less important features and preventing overfitting.
\item {\sf Feature Selection and Engineering:} Use techniques to reduce the feature set to those that are most relevant, eliminating noise from irrelevant features.
\item {\sf Ensemble Averaging:} Use techniques like bagging or averaging across multiple models or trees in the GBDT to reduce variance and stabilize predictions.
\item {\sf Pruning Trees and Early Stopping:} Regularly prune trees in the GBDT to prevent them from becoming too deep or too many and specific to the training data, thereby reducing noise sensitivity.
\item {\sf Robust Training:} Introduce noise or perturbations into the training data intentionally (e.g., via adversarial training) to help the model learn more robust decision boundaries.
\end{enumerate}

By identifying these factors and applying targeted strategies, you can significantly improve the robustness and stability of machine learning models, particularly in noisy environments. See \cite{cui2023enhancing} for further details about enhancing robustness of gradient-boosted decision trees. 

\subsection{Resilience Against Distribution Drift and Environment Change}
{\bf Resilience} refers to the model's ability to maintain accurate performance in the face of changes in the input data distribution or external factors. In banking, resilience is critical because economic conditions, customer behaviors, and regulatory environments can shift over time, impacting how well models perform.

{\bf Distribution Drift Analysis:} Over time, the distribution of input data (e.g., income levels, employment rates, customer behaviors) may shift, causing the model’s predictions to become less accurate. 
\begin{itemize}
\item {\sf Time-Based Analysis:} Evaluate model performance on different time slices of data to check for stability. This helps identify whether the model's effectiveness diminishes over time, indicating potential drift or the need for updates.
\item {\sf Segment Analysis:} Examine how well the model performs across various behavioral segments or clusters. Significant variations in performance across segments/clusters may suggest the need for model adjustments or the introduction of segment-specific models.
\item {\sf Stress Testing for Stability:} Perform stress tests by simulating extreme conditions (e.g., economic downturns) to evaluate how model predictions behave under stress. Stability under various stress conditions is crucial for ensuring the model's reliability in real-world scenarios.
\end{itemize}

{\bf Identification of Important Variables:} To assess resilience, it is crucial to identify which variables are most significant for the model's predictions and where changes in their distributions could significantly impact model performance \citep{sudjianto2023piml}. This involves:
\begin{itemize}
\item {\sf Information-Theoretic Measures:} Use information-theoretic metrics to assess feature importance by quantifying the distance between distributions of the features. Measures such as Jensen-Shannon Divergence (JSD) or Wasserstein Distance (see also {\bf Section 4.2.1}) can be employed to determine how much information a feature provides about the target variable. For instance:
\begin{itemize}
\item {\bf Jensen-Shannon Divergence} (also known as Population Stability Index/PSI) is a symmetric measure that quantifies the similarity between two probability distributions. A higher JSD indicates that changes in the feature distribution may significantly impact the model’s performance.
\item {\bf Wasserstein Distance}, also known as the Earth Mover's Distance, measures the cost of transforming one distribution into another. It provides a meaningful way to assess how distributions differ in terms of their shapes and support, capturing both the location and the spread of the distributions.
\end{itemize}

\item {\sf Monitoring Distribution Changes:} Once important variables are identified, their distributions should be monitored over time. For instance, if a model relies heavily on income and credit history, any significant shifts in the distribution of these features (e.g., a sudden increase in low-income borrowers due to economic conditions) could impact the model's ability to accurately assess credit risk.

\item {\sf Thresholds for Drift Detection:} Establishing thresholds for acceptable levels of drift in key variables can help in proactive monitoring. If the distance between the feature's current distribution and its historical distribution (as determined by Jensen-Shannon Divergence or Wasserstein Distance) exceeds a defined threshold, it can trigger a review of the model to determine if recalibration or retraining is necessary.

\item {\sf Using Information for Model Improvement:} The insights gained from variable distributions and their importance can also inform model improvement strategies. This includes:
\begin{itemize}
\item {\sf Feature Engineering:} Understanding how the distributions of key variables change can lead to new feature engineering opportunities. For example, if the income distribution shifts, creating interaction terms between income and other relevant features (like debt-to-income ratio) could enhance the model's predictive power.
\item {\sf Model Refinement:} Insights about important variables and their relationships with the target variable can guide model adjustments. For example, if certain variables show increasing divergence in their distributions, the model might require retraining or recalibration to accommodate these changes.
\item {\sf Segment-Specific Modeling:} If analysis reveals that certain segments (e.g., low-income borrowers) are particularly sensitive to distribution shifts, banks may choose to develop segment-specific models that better capture the unique behaviors and risks associated with those groups. This targeted approach can enhance overall model accuracy and robustness. To enhance resilience and ensure performance uniformity across different segments, employing a {\bf Mixture of Experts (MoE)} model can be beneficial. MoE models consist of multiple expert sub-models, each specializing in different regions of the input space or segments of the population. By dynamically routing inputs to the most appropriate expert model based on the context of the input features, the overall system can provide tailored predictions that account for varying behaviors across segments.

\medskip
{\bf Benefits of Mixture of Experts:} This approach allows for improved performance by ensuring that specific sub-models can better capture the nuances of particular segments (e.g., low-income borrowers, high-net-worth individuals) while the overarching model retains the ability to generalize across the entire population. MoE models can also enhance adaptability, as individual experts can be retrained or updated based on changes in the data distribution for their specific segments, maintaining accuracy and reducing the risk of underfitting or overfitting.

\item {\sf Dynamic Model Updates:} Leveraging insights from distribution changes can inform when to trigger model updates. By establishing clear criteria based on variable importance and distribution shifts, banks can implement more agile model management practices, allowing for timely adjustments in response to evolving data patterns.
\end{itemize}

\item {\sf Environmental Change Adaptation:} Resilient models can adapt to changes in the environment. Economic conditions, regulatory changes, or shifts in customer behavior can significantly alter the relationships between input variables and the target outcome. Regularly evaluating the model against different scenarios helps ensure that it remains relevant. For example, if interest rates change dramatically, the relationship between loan approval rates and income may shift, necessitating a review of the model's assumptions and structure.

\item {\sf Adaptive Maintenance:}To enhance resilience, banks should implement adaptive maintenance strategies that allow for timely updates to the model. This includes:
\begin{itemize}
\item {\sf Regular Recalibration:} Adjusting model parameters based on new data to reflect recent trends and changes in input distributions.
\item {\sf Model Refresh:} Periodically retraining the model with updated datasets to ensure it captures the latest customer behaviors and economic conditions.
\item {\sf Continuous Learning:} Incorporating mechanisms for the model to learn from new data continuously, ensuring it remains responsive to changes in the underlying patterns.
\end{itemize}
\end{itemize}

\section{Ongoing Monitoring}
{\bf Ongoing monitoring} is the final and continuous phase in the model validation process, aimed at ensuring the long-term performance and reliability of models after deployment. In banking, this phase is crucial because models operate in dynamic environments where data distributions, economic conditions, and regulatory requirements can change over time. Without proper monitoring, a model that initially performs well may degrade, leading to inaccurate predictions, increased risk, or regulatory non-compliance. The purpose of ongoing monitoring is to track model performance, detect issues early, and initiate corrective actions when necessary.

The core components of ongoing monitoring include continuous performance evaluation, testing for data drift, and periodic testing and revalidation to ensure the model remains accurate, robust, and compliant over time.

\subsection{Periodic Performance Monitoring}
Once a model is deployed in a production environment, periodic performance monitoring is essential to detect any deviations from expected behavior. This involves tracking key performance metrics in real time or at regular intervals to ensure the model is functioning correctly.
\begin{itemize}
\item {\sf Performance Tracking:} Banks typically monitor metrics such as accuracy, precision, recall, and area under the curve (AUC) for classification models. These metrics are tracked over time to detect any degradation in performance. For example, if a credit risk model begins to produce a higher false positive rate (predicting customers will default when they do not), it could indicate that the model is no longer reliable.
\item {\sf Error Analysis:} Regular error analysis helps to identify if specific types of errors, such as false positives or false negatives, are increasing. This is critical for high-stakes decisions, such as loan approvals, where misclassification can lead to financial losses or customer dissatisfaction.
\item {\sf Monitoring of Key Features:} In addition to output monitoring, it is important to track the behavior of key input features. For example, in a credit scoring model, monitoring the distribution of features like credit score, income, or debt-to-income ratio can help identify if the inputs are changing in ways that could affect model performance.
\end{itemize}

\subsection{Data Drift and Concept Drift Detection}
A primary focus of ongoing monitoring is the detection of {\bf data drift} (changes in the distribution of input data) and {\bf concept drift} (changes in the relationship between inputs and outputs) \citep{webb2016conceptdrift}. In banking, these drifts are particularly important because economic conditions, customer behaviors, and regulatory requirements can shift over time.


\subsubsection{Distribution Shift Detection}
Distribution shift refers to changes in the underlying probability distribution of data over time. These shifts can significantly affect the performance of machine learning models, making it crucial to detect them effectively. In this section, we provide formal mathematical formulations for various univariate and multivariate methods for detecting distribution shift, focusing on CDF-based and PDF-based approaches. We also discuss multivariate reconstruction error methods for identifying shifts in more complex datasets.

\subsubsection*{Univariate Distribution Shift Detection}
\begin{enumerate}
\item {\sf Kolmogorov-Smirnov (KS) Test.} 
The Kolmogorov-Smirnov test compares the empirical cumulative distribution functions (CDFs) of two distributions \( F(x) \) and \( G(x) \) based on the maximum distance between them.
\[
D_{n,m} = \sup_x | F_n(x) - G_m(x) |
\]
where \( F_n(x) \) and \( G_m(x) \) are the empirical CDFs of the two samples of size \( n \) and \( m \), respectively. \( D_{n,m} \) represents the KS statistic, and larger values indicate greater divergence between the two distributions.

\item {\sf Jensen-Shannon Divergence (JSD).} 
The Jensen-Shannon divergence (JSD) measures the similarity between two probability density functions (PDFs), \( P \) and \( Q \), and is a symmetric and smoothed version of the Kullback-Leibler (KL) divergence.
\[
\text{JSD}(P \parallel Q) = \frac{1}{2} \left( \text{KL}(P \parallel M) + \text{KL}(Q \parallel M) \right)
\]
where \( M = \frac{1}{2} (P + Q) \) and
\[
\text{KL}(P \parallel Q) = \sum_x P(x) \log \frac{P(x)}{Q(x)}
\]
JSD is bounded between 0 and 1, with 0 indicating identical distributions.

\item {\sf Kullback-Leibler Divergence (KL Divergence).}
KL divergence quantifies the difference between a reference probability distribution \( P(x) \) and an approximate distribution \( Q(x) \).
\[
\text{KL}(P \parallel Q) = \sum_x P(x) \log \frac{P(x)}{Q(x)}
\]
KL divergence is asymmetric, meaning \( \text{KL}(P \parallel Q) \neq \text{KL}(Q \parallel P) \), and is unbounded, where larger values indicate greater divergence.

\item {\sf Wasserstein Distance (Earth Mover's Distance).}
The Wasserstein distance measures the minimum "cost" of transforming one distribution into another by redistributing its probability mass. For univariate distributions, the first-order Wasserstein distance is given by:
\[
W_1(P, Q) = \int_{-\infty}^{\infty} | F_P(x) - F_Q(x) | dx
\]
where \( F_P(x) \) and \( F_Q(x) \) are the CDFs of distributions \( P \) and \( Q \), respectively.

\item {\sf Total Variation Distance (TVD).}
The Total Variation Distance measures the maximum difference in probabilities assigned by two distributions.
\[
\text{TVD}(P, Q) = \frac{1}{2} \sum_x | P(x) - Q(x) |
\]
It is bounded between 0 and 1, with 0 indicating that the two distributions are identical.
\end{enumerate} 

\subsubsection*{Multivariate Distribution Shift Detection}
\begin{enumerate}
\item {\sf Energy Distance.} 
Energy distance quantifies the difference between two multivariate distributions by comparing their energy representations. For two distributions \( P \) and \( Q \) with random variables \( X \) and \( Y \), respectively, and an independent copy \( X' \) of \( X \), the energy distance is given by:
\[
\mathcal{E}(P, Q) = 2 \mathbb{E}[\|X - Y\|] - \mathbb{E}[\|X - X'\|] - \mathbb{E}[\|Y - Y'\|]
\]
where \( \|\cdot\| \) denotes the Euclidean norm. This formulation captures both the means and variances of the distributions.

\item {\sf Maximum Mean Discrepancy (MMD).}
MMD measures the difference between the means of distributions in a reproducing kernel Hilbert space (RKHS). For two distributions \( P \) and \( Q \), the MMD is defined as:
\[
\text{MMD}^2(P, Q) = \mathbb{E}_P[k(X, X')] + \mathbb{E}_Q[k(Y, Y')] - 2 \mathbb{E}_{P, Q}[k(X, Y)]
\]
where \( k \) is a positive-definite kernel, such as the Gaussian or linear kernel. MMD is particularly useful for comparing high-dimensional distributions.

\item {\sf Reconstruction Error-based Approach.} This approach evaluates how well the data can be reconstructed. Two reconstruction methods can be employed: 
\begin{itemize}
    \item PCA that projects data $X$ onto principal components and low-dimensional representation \( \hat{X} \). 
    \item An autoencoder that is a neural network compressing data into a lower-dimensional representation and then reconstructing it to be  \( \hat{X} \).
\end{itemize}

In both methods, the reconstruction error is calculated as:
\[
\text{Reconstruction Error} = \| X - \hat{X} \|_2^2.
\]
A higher reconstruction error signifies that the data no longer conforms to the distribution on which the model was trained.

\item {\sf Mahalanobis Distance.}
Mahalanobis distance accounts for correlations between variables when measuring the distance between a point \( x \) and a distribution with mean \( \mu \) and covariance matrix \( \Sigma \):
\[
D_M(x, \mu, \Sigma) = \sqrt{(x - \mu)^\top \Sigma^{-1} (x - \mu)}
\]
This measure is particularly useful for detecting distribution shifts in multivariate datasets, as it adjusts for variance in each direction of the distribution.
\end{enumerate}

Detecting distribution shift is crucial for maintaining the performance of models, especially in dynamic environments where data distributions may change. The choice of univariate or multivariate methods depends on the complexity of the data and the type of shift being measured. While univariate methods like the Kolmogorov-Smirnov test and Jensen-Shannon divergence are effective for simpler data, multivariate methods such as energy distance, MMD, and reconstruction error-based approaches are essential for capturing more complex shifts in high-dimensional data.

\subsubsection{Concept Drift Detection}
Detecting concept drift, which occurs when the relationship between input features and output targets changes over time, is critical for ensuring the ongoing accuracy of predictive models. To rigorously diagnose concept drift, especially in the presence of potential input drift (covariate shift), we propose several methodologies that focus on isolating shifts in the input-output mapping. This section outlines various techniques to detect concept drift, including distribution comparison, residual analysis, and nearest-neighbor-based methods.

\begin{enumerate}
    \item {\sf Input Distribution Control via Nearest Neighbor Matching.} To distinguish concept drift from input drift, we propose a method that controls for changes in input distribution by matching new data to the training, validation, or test data using a nearest-neighbor approach. By identifying development data samples that are similar to new data points in the feature space, we can create a comparable input distribution across the two sets. This allows for a focused examination of how the model's predictive performance may have changed due to concept drift.

\begin{itemize}
    \item  Step 1: Nearest-neighbor selection. For each new data point, find its nearest neighbors in the development data (validation or test set) using distance metrics such as Euclidean, Mahalanobis, or embedding-based similarity. This ensures the selected subset of development data has a distribution similar to the new data.
    \item Step 2: Residual comparison. After controlling for input distribution, compute the residuals (i.e., the differences between actual and predicted outputs) for both the new data and the matched development data. Comparing the distribution of these residuals allows us to detect changes in the underlying input-output relationship. We apply statistical tests such as the Kolmogorov-Smirnov (KS) test, Cramér–von Mises test, or Anderson-Darling test to quantify whether the residual distributions from the new data and nearest-neighbor development data differ significantly. This allows us to rigorously test for concept drift.
\end{itemize}

With similar residual distributions, it suggests no concept drift. Performance degradation is likely due to input drift, as the model continues to capture the input-output relationship correctly but is faced with out-of-distribution inputs.

With divergent residual distributions: it indicates concept drift, where the model is no longer correctly modeling the relationship between inputs and outputs, even when controlling for input distribution.

\item {\sf Performance Monitoring Over Time (after Input Drift is Ruled Out).}  Once input drift has been ruled out using nearest-neighbor matching or similar methods, another robust indicator of concept drift is the degradation in model performance metrics (e.g., accuracy, precision, recall, F1-score, or RMSE) over time. Monitoring these metrics on a held-out test set or using rolling windows of time-based data enables the detection of gradual or sudden shifts in the input-output relationship.

The method of sliding window evaluation is to evaluate the model on successive time windows of new data, so we can track the stability of performance metrics. A consistent decline in performance across windows suggests that the model is encountering drift, either in the input space (input drift) or the input-output relationship (concept drift). However, if input drift has been ruled out via nearest-neighbor matching or other methods, the decline can be attributed to concept drift.

\item {\sf Residual Drift Monitoring (after Input Drift is Ruled Out).} Residuals provide a direct indication of how well the model captures the true relationship between inputs and outputs. After ruling out input drift using nearest-neighbor matching or other methods, monitoring the residuals over time allows for the detection of concept drift. Shifts in residual distributions—such as increased variance, systematic bias, or patterns in specific regions of the input space—can signal that the model is no longer aligned with the underlying data-generating process.

The method of localized drift detection is to account for localized concept drift (i.e., drift that occurs only in certain regions of the input space) by segmenting the input data based on feature importance or cluster similar input data points. By examining the residuals in each segment, we can detect localized shifts in the input-output mapping, which may not be evident when analyzing the entire dataset.

\item {\sf Model Retraining and Comparison.}  Retraining the model on more recent data and comparing its performance to the original model can also reveal concept drift. A significant improvement in the performance of the retrained model on recent data—compared to the original model—indicates that the original model no longer captures the current data distribution, suggesting concept drift.

We propose using paired statistical tests (e.g., paired t-tests) to rigorously compare the performance of the original model and the retrained model on the same test set. A statistically significant difference in predictions or error rates between the models indicates a change in the input-output relationship.

\item {\sf Model Error Tracking in Segments.} Segmenting data based on key features and tracking error rates in these segments helps identify regions in the input space where concept drift is more pronounced. By continuously monitoring model performance and residuals across different segments of data, we can pinpoint specific areas where the input-output relationship has shifted, enabling early detection of drift in localized contexts.
\end{enumerate}

These concept drift detection methods provide a rigorous framework for detecting concept drift by isolating changes in the input-output relationship. By leveraging nearest-neighbor-based distribution control with validation or test data, residual analysis, performance monitoring, and retraining comparisons, we can confidently distinguish between input drift and concept drift. This allows for timely model updates and ensures continued model reliability in dynamic environments.

\subsection{Periodic Testing and Revalidation}
{\bf Periodic testing} and {\bf revalidation} are essential components of ongoing monitoring to ensure that models remain fit for purpose over time. This involves formal, scheduled reviews of the model's performance and may be triggered either by routine cycles (e.g., quarterly, annually) or by specific events, such as significant economic changes or regulatory updates.
\begin{itemize}
\item {\sf Scheduled Periodic Testing:} Banks typically establish regular intervals (e.g., quarterly or annually) to conduct periodic testing. This involves running the model on fresh data that may not have been used in previous validations. The purpose is to assess whether the model continues to meet performance standards in light of recent data and economic conditions. For example, a credit scoring model might be tested on data from the most recent quarter to ensure it still accurately predicts defaults.
\item {\sf Revalidation in Response to Trigger Events:} In some cases, revalidation may be triggered by specific events, such as new regulations, significant market changes, or the discovery of performance issues during routine monitoring. For example, if the Federal Reserve changes interest rates, banks may need to revalidate risk models to account for how these changes impact customer loan repayment behavior.
\item {\sf Revalidation Process:} Revalidation involves a thorough review of the model, including retraining on new data, re-assessing assumptions, recalibrating parameters, and re-evaluating performance metrics. The process also includes stress testing the model under various economic scenarios to ensure that it remains robust and resilient. This revalidation may involve performing new rounds of backtesting, conducting cross-validation with updated datasets, and comparing the model’s performance against any regulatory benchmarks.
\item {\sf Documentation and Regulatory Reporting:} During periodic testing and revalidation, banks must ensure that the process is well-documented. This includes maintaining records of any changes to the model, rationale for those changes, and evidence that the model continues to meet regulatory and business standards. In the context of regulatory guidance, such as SR 11-7, banks are required to periodically report the results of their model validation efforts to regulators.
\end{itemize}

\subsection{Adaptive Maintenance and Model Refresh}
As part of ongoing monitoring, models may need to undergo adaptive maintenance, which includes updating model parameters or retraining the model with new data to reflect recent trends. This ensures that the model remains relevant in evolving environments.
\begin{itemize}
\item {\sf Model Recalibration:} Recalibration involves adjusting model parameters based on new data or changes in the external environment. For example, in a loan approval model, recalibration might be necessary if customers' repayment behavior changes due to macroeconomic factors like rising unemployment rates or inflation.
\item {\sf Model Retraining:} In some cases, recalibration may not be sufficient, and the model may need to be retrained from scratch. This typically happens when the model has been operating for an extended period, and significant data drift or concept drift has occurred. During retraining, new datasets, including recent customer behavior or updated economic conditions, are used to rebuild the model to ensure it remains effective and accurate.
\item {\sf Adaptive Models:} Some models, particularly in machine learning, are designed to be adaptive and can continuously learn from new data. These online learning models update their parameters in real-time as new data becomes available, ensuring that they stay current with the latest trends and conditions. However, the use of adaptive models requires careful monitoring to ensure that they do not inadvertently overfit to short-term noise or anomalies.
\end{itemize}

\section{Conclusion}
Model validation in banking is a complex, multifaceted process that plays a crucial role in ensuring the reliability, accuracy, and regulatory compliance of models. As we have explored in this paper, effective model validation encompasses three key components: conceptual soundness evaluation, outcome analysis, and ongoing monitoring.

The conceptual soundness evaluation forms the foundation of model validation, ensuring that models are built on solid theoretical grounds, use high-quality data, and align with established financial principles. This stage is critical in identifying potential issues early in the model development process, thereby reducing the risk of fundamental flaws in model design.

Outcome analysis provides a rigorous framework for assessing model performance, identifying weaknesses, and evaluating robustness. Through techniques such as performance decomposition, error analysis, and stress testing, banks can gain a deep understanding of how their models behave under various conditions. This understanding is crucial for building confidence in model outputs and for identifying areas where models may need refinement or additional controls.

The ongoing monitoring component recognizes that model validation is not a one-time event but a continuous process. In the dynamic world of finance, where economic conditions, customer behaviors, and regulatory requirements are constantly evolving, continuous vigilance is essential. Techniques for detecting data drift, concept drift, and changes in model performance help ensure that models remain effective and relevant over time.

As banking continues to evolve in the digital age, with increasing reliance on advanced analytics and machine learning, the importance of robust model validation practices will only grow. The challenges of maintaining model interpretability and managing the complexities of interconnected model ecosystems will require ongoing innovation in validation techniques.

The principles for model validation discussed here are also applicable to validation of more complex models including Generative AI --- though the detail testing approaches will be different. While not discussed here, there are other critical aspects including fairness \citep{zhou2021fairness}, model safety, security etc. that require important attention and will be a subject of a separate paper. 

In conclusion, effective model validation is essential for maintaining the integrity of banking operations, ensuring regulatory compliance, and building trust with customers and stakeholders. By embracing comprehensive validation practices that encompass conceptual soundness, rigorous outcome analysis, and diligent ongoing monitoring, banks can harness the power of predictive models while effectively managing the associated risks. As the financial landscape continues to evolve, so too must the practices of model validation, adapting to new challenges and opportunities in the pursuit of robust, reliable, and responsible banking models.


\begin{thebibliography}{}
\bibitem[Apley and Zhu, 2020]{apley2020visualizing}
Apley, D.~W. and Zhu, J. (2020).
\newblock Visualizing the effects of predictor variables in black box
  supervised learning models.
\newblock {\em Journal of the Royal Statistical Society Series B: Statistical
  Methodology}, 82(4):1059--1086.

\bibitem[Breunig et~al., 2000]{breunig2000lof}
Breunig, M. M., Kriegel, H. P., Ng, R. T., and Sander, J. (2000). LOF: identifying density-based local outliers. In {\em Proceedings of the 2000 ACM SIGMOD International Conference on Management of Data} (pp. 93-104).

\bibitem[Chen and Guestrin, 2016]{chen2016xgboost}
Chen, T, and Guestrin, C. (2016). 
Xgboost: A scalable tree boosting system.
{\em Proceedings of the 22nd ACM SIGKDD International Conference on Knowledge Discovery and Data Mining} (pp. 785--794). 

\bibitem[CFPB, 2022]{CFPB2022-03}
Consumer Financial Protection Bureau (2022). Consumer Financial Protection Circular 2022-03: Adverse action notification requirements in connection with credit decisions based on complex algorithms. May 26, 2022. 

\bibitem[Cui et~al., 2023]{cui2023enhancing}
Cui, S., Sudjianto, A., Zhang, A., and Li, R. (2023).
\newblock Enhancing robustness of gradient-boosted decision trees through
  one-hot encoding and regularization.
\newblock {\em arXiv preprint:2304.13761}.

\bibitem[Federal Reserve, 2011]{SR11-7}
Federal Reserve (2011). {\em Supervisory Guidance on Model Risk Management}. Board of Governors of the Federal Reserve System: SR Letter 11--7, Office of the Comptroller of the Currency 2011-12.  

\bibitem[Friedman(2001)]{friedman2001greedy}
Friedman, J. H. (2021)
\newblock Greedy function approximation: a gradient boosting machine.
\newblock \emph{Annals of Statistics}, 29(5):\penalty0 1189--1232.

\bibitem[Hastie et~al., 2015]{hastie2015statistical}
Hastie, T., Tibshirani, R., and Wainwright, M. (2015).
\newblock {\em Statistical Learning with Sparsity: the Lasso and Generalizations}.
\newblock CRC press.

\bibitem[Hu et~al., 2023]{hu2023interpretable}
Hu, L., Nair, V.~N., Sudjianto, A., Zhang, A., and Chen, J. (2023).
\newblock Interpretable machine learning based on functional ANOVA framework: algorithms and comparisons.
\newblock {\em arXiv preprint:2305.15670}.

\bibitem[Ke et~al., 2017]{ke2017lightgbm}
Ke, G., Meng, Q., Finley, T., Wang, T., Chen, W., Ma, W., Ye, Q. and Liu, T. Y. (2017). Lightgbm: A highly efficient gradient boosting decision tree. 
{\em Advances in Neural Information Processing Systems}, 30.

\bibitem[Liu et~al., 2008]{liu2008isolation}
Liu, F. T., Ting, K. M., and Zhou, Z. H. (2008). Isolation forest. In {\em 2008 Eighth IEEE International Conference on Data Mining} (pp. 413-422). 

\bibitem[Lou et~al., 2013]{lou2013accurate}
Lou, Y., Caruana, R., Gehrke, J., and Hooker, G. (2013).
\newblock Accurate intelligible models with pairwise interactions.
\newblock In {\em Proceedings of the 19th ACM SIGKDD International Conference
  on Knowledge Discovery and Data Mining} (pp. 623--631). ACM.

\bibitem[Lundberg and Lee, 2017]{lundberg2017shap}
Lundberg, S.~M. and Lee, S.-I. (2017).
\newblock A unified approach to interpreting model predictions.
\newblock {\em Advances in Neural Information Processing Systems}, 30.





\bibitem[Ribeiro et~al., 2016]{ribeiro2016lime}
Ribeiro, M.~T., Singh, S., and Guestrin, C. (2016).
\newblock ``Why should i trust you?'' {E}xplaining the predictions of any
  classifier.
\newblock In {\em Proceedings of the 22nd ACM SIGKDD international conference
  on knowledge discovery and data mining} (pp. 1135--1144).


\bibitem[Schubert et~al., 2017]{schubert2017dbscan}
Schubert, E., Sander, J., Ester, M., Kriegel, H. P., and Xu, X. (2017). DBSCAN revisited, revisited: why and how you should (still) use DBSCAN. {\em ACM Transactions on Database Systems}, 42(3), 1--21.

\bibitem[Shafer and Vovk, 2008]{shafer2008conformal}
Shafer, G., and Vovk, V. (2008). A tutorial on conformal prediction. {\em Journal of Machine Learning Research}, 9(3):371--421.

\bibitem[Sudjianto et~al., 2020]{sudjianto2020unwrapping}
Sudjianto, A., Knauth, W., Singh, R., Yang, Z., and Zhang, A. (2020).
\newblock Unwrapping the black box of deep relu networks: interpretability,
  diagnostics, and simplification.
\newblock {\em arXiv preprint:2011.04041}.

\bibitem[Sudjianto and Zhang, 2021]{sudjianto2021design}
Sudjianto, A. and Zhang, A. (2021). 
Designing inherently interpretable machine learning models. 
{\em arXiv preprint:2111.01743}. 


\bibitem[Sudjianto et~al., 2023]{sudjianto2023piml}
Sudjianto, A., Zhang, A., Yang, Z., Su, Y., and Zeng, N. (2023). PiML toolbox for interpretable machine learning model development and diagnostics. 
{\em arXiv preprint:2305.04214.}

\bibitem[Vovk et~al., 2005]{vovk2005}
Vovk, V., Gammerman, A., and Shafer, G. (2005). {\em Algorithmic Learning in a Random World.} Springer, New York.

\bibitem[Webb et~al., 2016]{webb2016conceptdrift}
Webb, G. I., Hyde, R., Cao, H., Nguyen, H. L., and Petitjean, F. (2016). Characterizing concept drift. {\em Data Mining and Knowledge Discovery}, 30(4), 964-994.

\bibitem[Yang et~al., 2020]{yang2020enhancing}
Yang, Z., Zhang, A. and Sudjianto, A. (2020).
\newblock Enhancing explainability of neural networks through architecture
  constraints.
\newblock {\em IEEE Transactions on Neural Networks and Learning Systems},
  32(6):2610--2621.

\bibitem[Yang et~al., 2021]{yang2021gami}
Yang, Z., Zhang, A., and Sudjianto, A. (2021).
\newblock {GAMI-Net}: An explainable neural network based on generalized
  additive models with structured interactions.
\newblock {\em Pattern Recognition}, 120:108192.

\bibitem[Zhao et~al., 2019]{zhao2019pyod}
Zhao, Y., Nasrullah,  Z. and Li, Z. (2019). 
\newblock PyOD: A Python toolbox for scalable outlier detection
\newblock {\em Journal of Machine Learning Research}, 20(96):1--7. 


\bibitem[Zhou et~al., 2021]{zhou2021fairness}
Zhou N., Zhang Z., Nair V., Singhal H., Chen J. and Sudjianto A. (2021).
\newblock   Bias, fairness, and accountability with AI and ML algorithms.
\newblock {\em arXiv preprint:2105.06558}.

\end{thebibliography}
\end{document}